\documentclass{aastex63}
\usepackage{xcolor}
\usepackage{subfigure}
\usepackage{amsmath}
\usepackage{color, soul}
\setstcolor{red}

\received{2021 March 11}
\revised{2021 June 8}
\accepted{2021 June 9}
\submitjournal{ApJ}

\shorttitle{A necessary condition for SN fallback invading newborn NS magnetosphere}
\shortauthors{Zhong et al.}

\begin{document}
\title{A necessary condition for supernova fallback invading newborn neutron-star magnetosphere}

\correspondingauthor{Yici Zhong}
\email{yici.zhong@phys.s.u-tokyo.ac.jp}

\author{Yici Zhong}
\affiliation{Department of Physics, School of Science, The University of Tokyo, 7-3-1 Hongo, Bunkyo-ku, Tokyo 113-0033, Japan}

\author{Kazumi Kashiyama}
\affiliation{Research Center for the Early Universe (RESCEU), School of Science, The University of Tokyo, 7-3-1 Hongo, Bunkyo-ku, Tokyo 113-0033, Japan}
\affiliation{Department of Physics, School of Science, The University of Tokyo, 7-3-1 Hongo, Bunkyo-ku, Tokyo 113-0033, Japan}

\author{Toshikazu Shigeyama}
\affiliation{Research Center for the Early Universe (RESCEU), School of Science, The University of Tokyo, 7-3-1 Hongo, Bunkyo-ku, Tokyo 113-0033, Japan}
\affiliation{Department of Astronomy, School of Science, The University of Tokyo, 7-3-1 Hongo, Bunkyo-ku, Tokyo 113-0033, Japan}

\author{Shinsuke Takasao}
\affiliation{Department of Earth and Space Science, Graduate School of Science, Osaka University, Toyonaka, Osaka 560-0043, Japan}

\begin{abstract}
We numerically investigate the dynamics of a supernova fallback accretion confronting with a relativistic wind from a newborn neutron star (NS). 
The time evolution of the accretion shock in the radial direction is basically characterized by the encounter radius of the flow $r_\mathrm{enc}$ and a dimensionless parameter $\zeta \equiv L/\dot M_\mathrm{fb}c^2$, where $L$ is the NS wind luminosity and $\dot M_\mathrm{fb}$ is the fallback mass accretion rate. We find that the critical condition for the fallback matter to reach near the NS surface can be simply described as $\zeta < \zeta_\mathrm{min} \equiv GM_*/c^2r_\mathrm{enc}$ or  $r_\mathrm{enc}L/G M_* \dot M_\mathrm{fb} < 1$ independent of the wind Lorentz factor, where $M_*$ is the NS mass. 
With combining the condition for the fallback matter to bury the surface magnetic field under the NS crust, we discuss the possibility that the trifurcation of NSs into rotation-powered pulsars, central compact objects (CCOs), and magnetars can be induced by supernova fallback. 
\end{abstract}


\section{Introduction}\label{sec:intro}

Young neutron stars (NSs) in the Galaxy with ages of $t_\mathrm{age} \lesssim 1\mbox{-}10$ kyr are categorized into three classes: (non-recycled) pulsars, magnetars, and central compact objects (CCOs)~\cite[e.g.,][for a review]{enoto2019observational}. Based on the multi-wavelength information, their main energy sources are considered to be different: rotation energy, magnetic field energy, and latent heat, respectively. One of the key parameters is the magnetic field strength; the strengths of the dipole field are estimated to be $B_\mathrm{d} \sim 10^{12\mbox{-}13}\,\mathrm{G}$ for rotation-powered pulsars, $B_\mathrm{d} \gtrsim 10^{14}\,\mathrm{G}$ for magnetars, and $B_\mathrm{d} \lesssim 10^{11}\,\mathrm{G}$ for CCOs. The origin of the diversity is still unsettled.  

The magnetic field strength of a young NS should be determined as a consequence of various processes.
Most of the NSs are formed in collapsing massive stars, where the magnetic field of the progenitor core can be amplified by the flux-freezing contraction~\citep{1964ApJ...140.1309W}, the $\alpha$-$\Omega$ dynamo~\citep[e.g.,][]{Duncan1992neutron,thompson1993neutron}, the magnetorotational instability~\citep[e.g.,][]{2003ApJ...584..954A,2005ApJ...620..861T} occurring in the proto-NS, and/or stationary accretion shock instability (SASI) of the post-bounce core-collapse supernova environment~\citep[e.g.,][]{2012ApJ...751...26E}. 
On the other hand, the magnetic field can also decay with a relatively long timescale via the combination of the ambipolar diffusion, the Hall drift, and the Ohmic diffusion~\citep{GR92}.

The supernova fallback has been also considered to be relevant, especially for explaining the apparently weak magnetic field of the CCOs~\citep[e.g.,][]{muslimov1995delayed,Torres_Forn__2016}. 
If the fallback accretion proceeds down to the near surface region, it disturbs the NS magnetosphere. In the extreme case, the fallback matter can bury the surface magnetic field down in the non-convective crust. In this scenario, the bifurcation between CCOs and other types of NS can be determined by the competition of the fallback accretion and the outflow from the newborn NS; if the outflow repulses the fallback matter, the central NS evolves into a pulsar, and otherwise a CCO with buried magnetic fields is formed~\citep{SK18}.

The competition between the outflow from the newborn NS and the fallback inflow will occur in the following manner. When a (proto-)NS is formed, a sub- or trans-relativistic neutrino-driven wind is initially the dominant outflow process~\mbox{\citep[e.g.,][]{vincenzo2021nucleosynthesis}}. The neutrino-driven wind is considered to last for $\sim$ 10 sec, which corresponds to the neutrino cooling timescale of the proto-NS. The neutrino driven wind catches up to the tail of the supernova ejecta and pushes it outward. During this period, a nascent magnetosphere will be formed inside the ``bubble" produced by the neutrino-driven wind. When the neutrino luminosity of the proto-NS decreases and the neutrino-driven wind ceases, a fraction of the tail of the supernova ejecta can start to fall back~\mbox{\citep[e.g.,][]{ugliano2012progenitor}}. At the same timing, the dominant outflow process will be switched to a relativistic rotation-powered wind~\mbox{\citep[e.g.,][]{Gruzinov05,Spitkovsky06,Tchekhovskoy+13}}, which confronts with the supernova fallback.


In order to derive the critical condition for the fallback matter to reach near the NS surface, \cite{SK18} constructed a self-similar solution for a spherically symmetric fallback accretion confronting with a relativistic outflow, which is a one-parameter family of the out- to inflow luminosity ratio. 
However, it is also important to investigate the impacts of other physical quantities,  
e.g., the encounter radius of the in- and outflows and the Lorentz factor of the outflow. To this end, we perform a suit of relativistic hydrodynamic simulations and clarify the condition for the fallback matter invading down to the NS magnetosphere. 

This paper is organized as follows.
We describe the problem setting in Sec. \ref{sec:setup}, and show the results of the numerical simulation in Sec. \ref{sec:result}. We consider the implications of the results for the diversity in young NSs in Sec. \ref{sec:implication}. Sec. \ref{sec:discussion} is devoted to the summary and discussion. 
We use the convention of $Q_x=Q/10^x$ in cgs units unless otherwise noted.

\section{Setup}\label{sec:setup}

In a successful core-collapse supernova explosion, a bulk of the progenitor mass becomes gravitationally unbound and is ejected, but a tail part of the ejecta can become marginally bound and falls back to the newborn NS. Such a mass accretion can be induced either when the neutrino luminosity from the NS significantly decreases~\mbox{\citep[e.g.,][]{ugliano2012progenitor}} or the supernova shock clashes into the thick outer envelope~\mbox{\citep{chevalier1989neutron}}. Here we mainly consider the former case. When the neutrino luminosity of the proto-NS decreases and the neutrino-driven wind ceases, a rotation-powered relativistic wind is the dominant outflow process.
Here we numerically investigate the dynamics of the supernova fallback accretion confronting with the relativistic wind in order to find the critical condition for the fallback matter to reach near the NS surface.


\begin{figure}[htp]
    \centering
    \includegraphics[width=14cm]{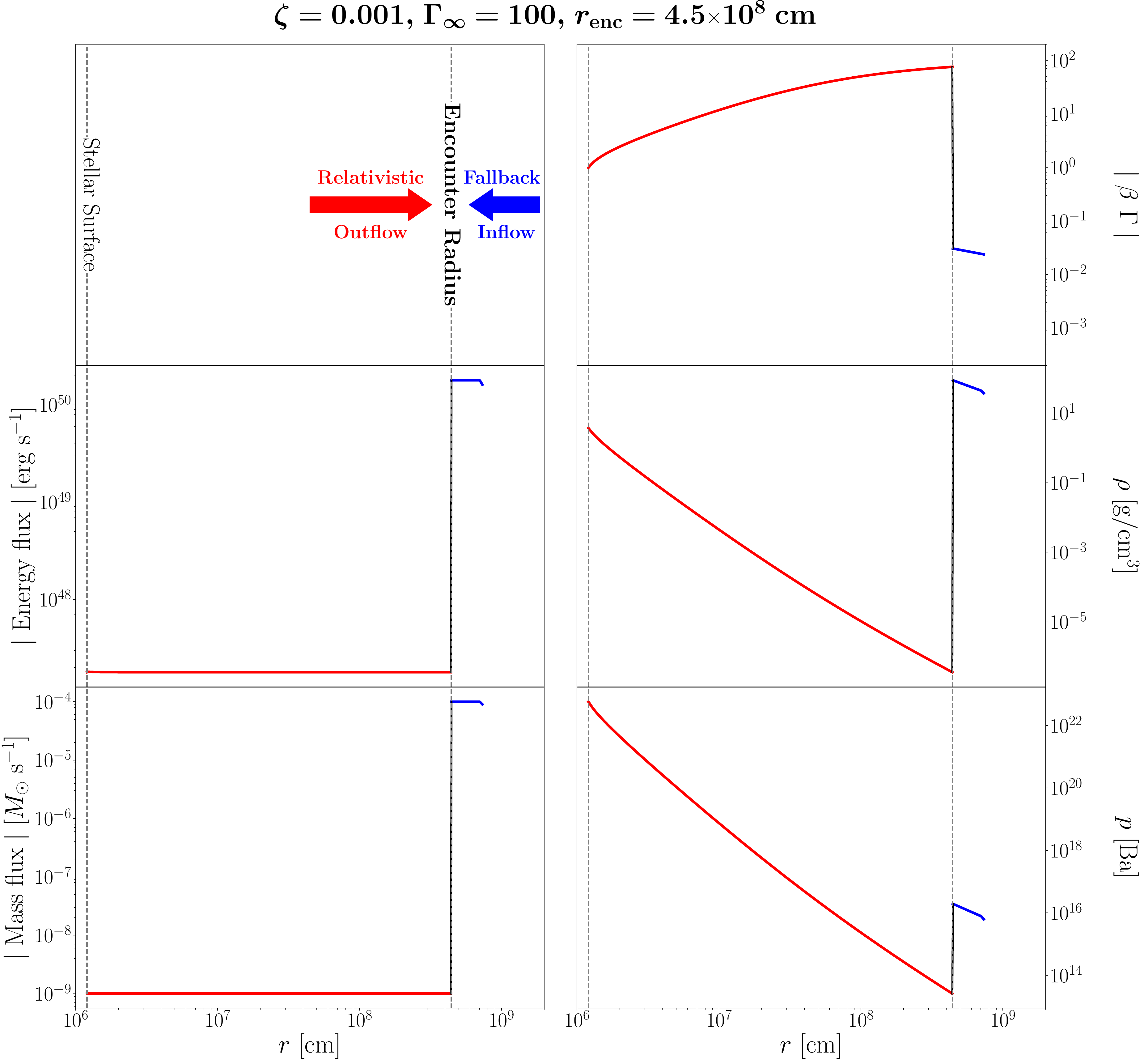}
    \caption{An example of initial condition of our simulation. A relativistic outflow with a terminal Lorentz factor $\Gamma_\infty = 100$ collides with a fallback matter at the encounter radius $r_\mathrm{enc} = 4.5\times 10^{8}\,\mathrm{cm}$. The dimensionless out- to inflow luminosity ratio (Eq. \ref{eq:zeta}) is $\zeta = 0.001$. As for the energy and mass fluxes and the velocity $\beta\Gamma$, the signs are positive in the outflow region and negative in the inflow region.
    }
    \label{fig:fall_back}
\end{figure}

\subsection{Initial conditions}

As shown in Fig. \ref{fig:fall_back}, we consider a fallback matter with a mass accretion rate $\dot M_\mathrm{fb}$ and a relativistic outflow with a luminosity of $L$ and a terminal Lorentz factor of $\Gamma_\infty$ to encounter at a radius of $r = r_\mathrm{enc}$ (a detailed description will be given in Sec. \mbox{\ref{xenc}}). The system is characterized by three dimensionless parameters ($\zeta$, {${\cal R}_\mathrm{enc}$}, $\Gamma_\infty$), where 
\begin{equation}\label{eq:zeta}
\zeta = \frac{L}{\dot M_\mathrm{fb,ini}c^2}
\end{equation}
is a dimensionless parameter representing the ratio between the outflow luminosity and the initial fallback accretion rate $\dot M_\mathrm{fb,ini}$ 
and
\begin{equation}\label{eq:Sch}
{\cal R}_\mathrm{enc} = \frac{r_\mathrm{enc}}{r_\mathrm{Sch}}
\end{equation}
with $r_\mathrm{Sch} = 2GM_*/c^2$ being the Schwarzschild radius of the central NS with mass $M_*$.
In this paper, we consider a spherically symmetric one-dimensional flow in order to explore a wide range of these parameters. 

\subsubsection{Fallback accretion}\label{fall-back}

The fallback accretion typically sets in at $t_{\rm fb} \sim 10\,{\rm s}$ after the explosion and the total fallback mass ranges over $M_\mathrm{fb} \sim 10^{-(2\mbox{-}4)}\,M_\odot$, depending on the core structure of the progenitor~\citep[e.g.,][]{ugliano2012progenitor,2016ApJ...821...69E}. Accordingly, the fallback accretion rate ranges over $\dot M_{\rm fb} \sim 10^{-3}-10^{-6}\,M_\odot\,{\rm s^{-1}}$.
We assume the fallback accretion rate as 
\begin{equation}\label{eq:fallbackrate}
    \dot M_\mathrm{fb} = \dot M_\mathrm{fb,ini} \times 
    \begin{cases} 
    1 & t \leq t_\mathrm{fb} \\
    (t/t_\mathrm{fb})^{-l} & t > t_\mathrm{fb}
    \end{cases},
\end{equation}
where 
\begin{equation}
\dot M_\mathrm{fb,ini} = \frac{l-1}{l}\frac{M_\mathrm{fb}}{t_\mathrm{fb}} \sim 1\times 10^{-5}\,M_\odot\,\mathrm{s}^{-1}\,\left(\frac{l-1}{l}\right)M_\mathrm{fb,-4}t_\mathrm{fb,1}{}^{-1},
\end{equation}
and $l > 1$ so that the total fallback mass is $M_\mathrm{fb}$.
Here $M_\mathrm{fb,-4} = M_\mathrm{fb}/10^{-4}\,M_\odot$.
We fix $l = 5/3$, which is expected for an accretion of marginally gravitationally bound matter\mbox{~\citep[e.g., ][]{chevalier1989neutron,2021arXiv210407493J}}.  The fallback matter is set to have the free-fall velocity at each radius $r$, 
\begin{equation} \label{eq:fallback_vel}
v_\mathrm{fb}(r)= - \sqrt{\frac{2 G M_*}{r}},
\end{equation}
where $M_* = 1.4 \, M_\odot$ is the neutron star mass.
The density profile is determined by assuming that the fallback matter is in a steady state with the inward mass flux given by Eq. (\ref{eq:fallbackrate}) and the velocity $v_\mathrm{fb}(r)$.
We assume that the unshocked fallback matter is sufficiently cold.

\subsubsection{Relativistic wind}\label{pulsarwind}
As a confronting outflow to the fallback accretion, we consider a relativistic wind powered by the spindwon luminosity of the newborn NS~\citep[e.g.,][]{pacini1967energy,1969ApJ...157.1395O}. 
The wind luminosity should be determined by the magnetic field strength and the angular frequency of the NS (see Sec. \ref{sec:implication}).
On the other hand, the Lorentz factor depends on the baryon loading and magnetization of the embryonic magnetosphere, which are highly uncertain. 
We here consider a relativistic hydrodynamic wind with a terminal Lorentz factor of $\Gamma_\infty$, and parametrically study the impact on the fallback accretion dynamics. 
In this case, the wind profile can be obtained by solving the following equations for a given set of $(L, \Gamma_\infty)$;
\begin{equation} \label{eq1}
4 \pi r^{2} \beta_\mathrm{w} \Gamma_\mathrm{w}{}^{2} \rho_\mathrm{w} h_\mathrm{w} = L,
\end{equation}
\begin{equation} \label{eq2}
4 \pi \Gamma_\mathrm{w} \beta_\mathrm{w} \rho_\mathrm{w} r^{2} = \frac{L}{\Gamma_{\infty}},
\end{equation}
where $\beta_\mathrm{w}$ is the velocity, $\rho_\mathrm{w}$ is the proper mass density, $\Gamma_\mathrm{w} = 1/\sqrt{1-\beta_\mathrm{w}{}^2}$ is the Lorentz factor, and $h_\mathrm{w} =1+\gamma/(\gamma-1) \times p_\mathrm{w}/\rho_\mathrm{w}$ is the specific enthalpy, and $p_\mathrm{w} = k_\mathrm{w} \rho_\mathrm{w}{}^{\gamma}$ is the pressure with $\gamma = 4/3$ being the adiabatic index. 
We assume a trans-relativistic wind velocity at the inner most radius, $\beta_\mathrm{w} (R_*) = 0.7$~\footnote{This treatment enhances the numerical stability; in the case of setting a relativistic velocity at the inner boundary, a numerical instability occurs when the reverse shock approaches the boundary.} and set the constant coefficient $k_\mathrm{w}$ so that the Lorentz factor of the wind becomes $\Gamma_\infty$ at infinity. Note that since the spindown timescale $t_\mathrm{sd}$ is typically much longer than the dynamical timescale we are interested in (see Eq. \ref{eq:tsd}), we assume that the wind luminosity is constant in the following calculations.

\subsubsection{The encounter radius}\label{xenc}

Since the fallback timescale should be roughly a free-fall timescale from the fallback radius ($r_\mathrm{fb}$), the fallback radius is given as
\begin{equation}
    r_{\rm fb} = (GM_* t_{\rm fb}{}^2)^{1/3} \sim 2.7\times 10^9\,{\rm cm}\,t_{\rm fb, 1}{}^{2/3}.
\end{equation}
In the case of a relativistic wind, the propagation timescale of the wind from the NS surface to the fallback radius is negligible compared with the fallback timescale. Thus, the relativistic wind and the fallback matter should encounter practically at
\begin{equation}\label{eq:r_enc}
    r_\mathrm{enc} \approx r_\mathrm{fb} \sim 2.7\times 10^9\,{\rm cm}\,t_{\rm fb, 1}{}^{2/3}.
\end{equation}
We note that $r_\mathrm{fb}$ and $r_\mathrm{enc}$ should be determined as a result of the complex supernova explosion dynamics and sensitive to the core structure of the progenitor star. We here define it as a model parameter of our simulation.
We also note that the encounter radius is typically much larger than the light cylinder radius,
\begin{equation}\label{r_lc}
    r_{\rm lc} = \frac{c P_\mathrm{i}}{2\pi} \sim 4.8\times 10^{7}\,{\rm cm}\,P_\mathrm{i,-2}{},
\end{equation}
where $P_\mathrm{i}$ is the initial spin period of the NS.

\subsection{Numerical simulation} \label{numerical_simu}
For the given initial condition in the previous section, the time evolution of the shock structure is obtained by numerically solving one-dimensional relativistic hydrodynamic equations with a gravity source term
under the spherical symmetry; 
\begin{equation}\label{eq:mass_con_grav}
\frac{\partial D}{\partial t}+\frac{1}{r^{2}} \frac{\partial}{\partial r}\left(D \beta r^{2}\right)=0,
\end{equation}
\begin{equation}\label{eq:momen_con_grav}
\frac{\partial S}{\partial t}+\frac{1}{r^{2}} \frac{\partial (r^{2}S \beta)}{\partial r} +\frac{\partial p}{\partial r}= -\frac{G M_*}{r^{2}} D.
\end{equation}
\begin{equation}\label{eq:ene_con_grav}
\frac{\partial E}{\partial t}+\frac{1}{r^{2}} \frac{\partial}{\partial r} \left(r^{2}S \right)= -\frac{G M_*}{r^{2}} S.
\end{equation}
Here $D = \Gamma \rho$, $S = \Gamma^{2} \rho h\beta$, and $E = \Gamma^{2}\rho h-p$ represent the mass, momentum, and energy densities, respectively. 
We assume the equation of state $h=1+\gamma p/\rho (\gamma-1)$ with a constant adiabatic index $\gamma = 4/3$. 
For a given fallback rate $\dot M_\mathrm{fb}$, we set the velocity and density at the outer boundary following Eq.(\ref{eq:fallback_vel}) and $\rho(r_\mathrm{out}) = \dot M_\mathrm{fb} / (4 \pi r^{2} v_\mathrm{fb}(r_\mathrm{out}))$. The pressure at the outer boundary is given by fixing the sound velocity as $c_\mathrm{s} \sim 10^{-3}c$. On the other hand, for a given outflow luminosity $L$, the density and pressure at the inner boundary are determined from Eqs.(\ref{eq1}) and (\ref{eq2}) and the equation of state by fixing the velocity as $\beta_\mathrm{w}(R_*) = 0.7$.

\if
Conservation laws under special relativity can be expressed as follows:
\begin{equation}\label{eq:mass_con}
\left(\rho u^{\mu}\right)_{, \mu}=0,
\end{equation}
\begin{equation}\label{eq:momen_ener_con}
T^{\mu \nu}{}_{, \nu}= F^{\mu} ,
\end{equation}
where $F^{\mu}$ stands for source term that can be specified regarding physical settings, $u^{\mu}$ and $T^{\mu \nu}$ represents the velocity four-vector and stress-momentum tensor, respectively:
\begin{equation}\label{eq:v_4_vector}
u^{v}=\Gamma\left(1, u_{i}, u_{j}, u_{k}\right), \Gamma=1/\sqrt{1-\beta^{2}},
\end{equation}
\begin{equation}\label{eq:E_P_tensor}
T^{\mu \nu}=\rho h u^{\mu} u^{\nu}+P \eta^{\mu \nu},
\end{equation}
\begin{equation}\label{eq:metric tensor}
\eta^{\mu \nu}=\left(\begin{array}{cccc}
-1 & 0 & 0 & 0 \\
0 & 1 & 0 & 0 \\
0 & 0 & 1 & 0 \\
0 & 0 & 0 & 1 
\end{array}\right),
\end{equation}
where $\rho$ is proper mass density, $P$ is gas pressure, velocity of light $c$ is taken as 1 here. $h$ stands for special enthalpy and can be expressed as:
\begin{equation}\label{eq:enthalpy}
h=\epsilon+\frac{P}{\rho},
\end{equation}
where $\epsilon$ is specific internal energy, and can be well-defined as long as the equation of state is fixed. Here we assume an adiabatic equation of state with an adiabatic index of $\gamma$, which is expected to be $4/3$ in ultra-relativistic cases. Under this setting, $h$ can be specified as:
\begin{equation}\label{eq:enthalpy_ourcase}
h=1+\frac{P}{(\gamma-1) \rho}+\frac{P}{\rho}.
\end{equation}
\fi

We use the Athena++ code \citep{2020ApJS..249....4S} for the numerical integration.
We employ the Harten-Lax-van Leer-Contact (HLLC) Riemann solver~\citep{10.1111/j.1365-2966.2005.09546.x} 
and use the second-order piecewise linear reconstruction method (PLM) with van Leer slope limiter~
\citep{VANLEER1974361}. 
The time integration is carried out by the second order Runge-Kutta method with a Courant-Friedrich-Lewy number of 0.1.
The inner boundary is fixed to be at $R_* = 12\,\mathrm{km}$. The radius of the outer boundary is set to be sufficiently larger than the encounter radius of the in- and outflows. 

The computational domain is resolved with the mesh number of 1024. We employ a non-uniform mesh, where the radial grid size is proportional to the radius. The fiducial value of the grid size ratio $\Delta r(i+1)/\Delta r(i) $ is 1.009.
The convergence of the numerical results with respect to the spatial resolution has been confirmed.
Given the mesh spacing, we set the initial profile shown in the previous sections with the cubic B spline in the outflow region.  

We investigate the range of the parameters shown in Table {\ref{table:1}. We choose four different encounter radii ($r_\mathrm{enc} = 4.5 \times 10^7$, $1.8 \times 10^8$, $4.5 \times 10^8$, $1.8\times 10^9$ cm) and three different terminal Lorentz factors ($\Gamma_\infty=6, 10$ and $100$). For a given combination of $(r_\mathrm{enc}/r_\mathrm{Sch}, \Gamma_\infty)$, we try a few 10 different $\zeta$ values in the range of [$10^{-5}:10$]. When varying $\zeta$, we fix the fallback accretion rate at $\dot{M}_\mathrm{fb,i}=10^{-4}\,M_\odot\,\mathrm{s^{-1}}$ and vary the wind luminosity $L$.}~\footnote{We have confirmed that the same ($\zeta$, ${\cal R}_\mathrm{enc}$, $\Gamma_\infty$) but different ($L$,$\dot M_\mathrm{fb ,i}$) cases give the same minimum fallback radii.} In total, we calculate $\sim 200$ cases.

\begin{table}[ht]
\centering
\begin{tabular}{ccc}
\hline \text { parameter } & \text { notation } & \text { range } \\
\hline \text { out- to inflow luminosity ratio } & {$\zeta$} & [$10^{-5}$ : $10$] \\\text { outflow Lorentz factor at infinity} &
{$\Gamma_{\infty}$} & (1 : 100] \\\text { normalized encounter radius}  &
{${\cal R}{\rm_{enc}}$} & [100 :  10000] \\
\hline
\end{tabular}
\caption{Run parameters}
\label{table:1}
\end{table}

\section{Results} \label{sec:result}

\begin{figure}[htp]
    \centering
    \includegraphics[width=14cm]{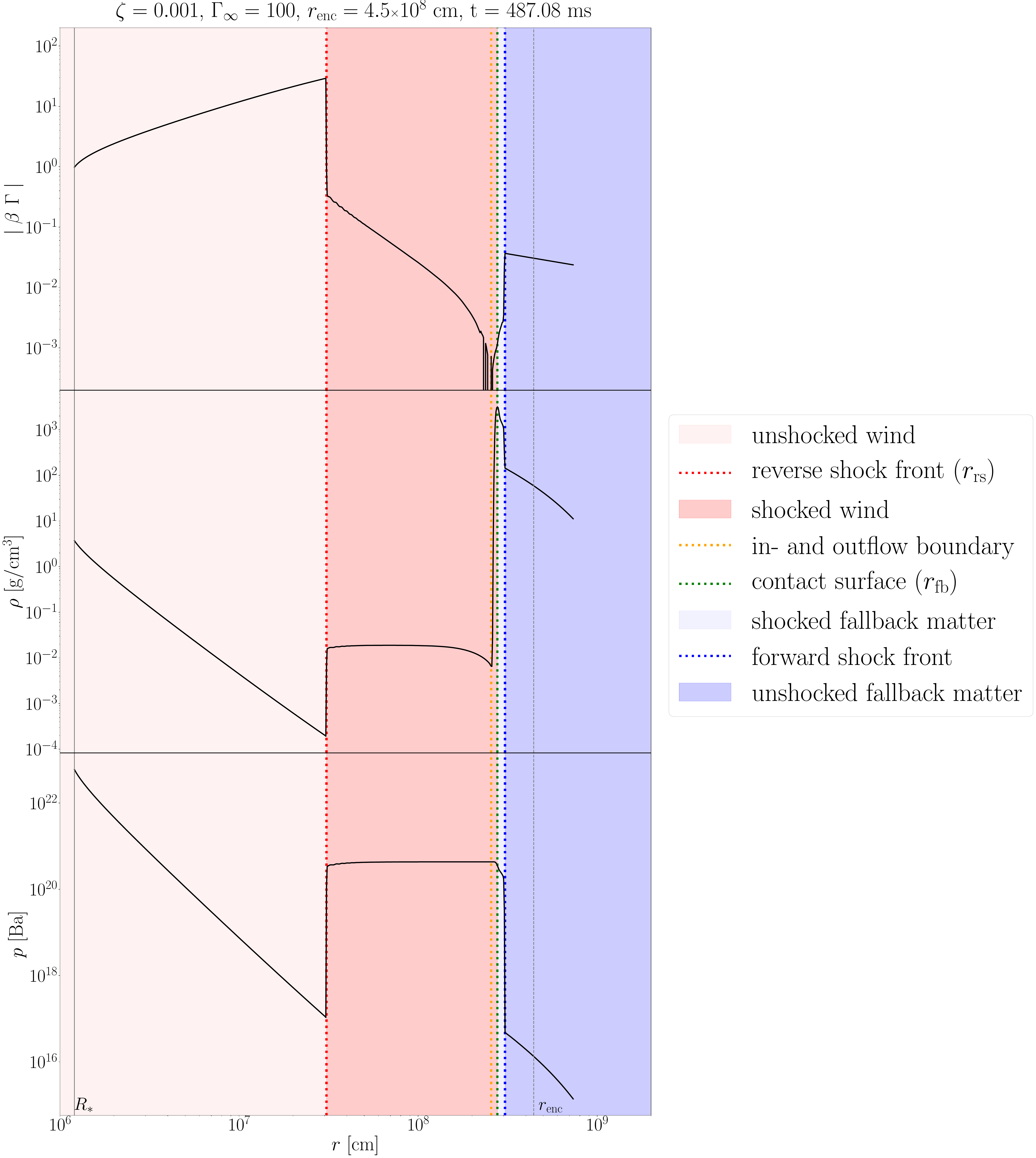}
    \caption{Shock structure formed between a fallback matter and a relativistic outflow with a terminal Lorentz factor of $\Gamma_\infty = 100$, encountered at a radius of $r_{\mathrm{enc}} = 4.5 \times 10^{8}$ cm (dashed line). We show the case with an out- to inflow luminosity ratio $\zeta=0.001$. The top, middle, and bottom panels show the velocity, density, and pressure profiles for $t=487.08\,\mathrm{ms}$, respectively. The discontinuities and the stellar surface are marked by the vertical dotted and solid lines, respectively.}
    \label{fig:shock structure}
\end{figure}

As shown in Fig. \ref{fig:shock structure}, multiple discontinuities form when an inflow and an outflow collide. The shocked and unshocked fallback matter are separated by a forward shock while the shocked and unshocked winds are separated by a reverse shock. The shocked fallback matter and the wind are separated by a contact surface, at which the gas density takes its maximum value.
In this sense, the position of contact surface $r_{\mathrm{fb}}$
can be regarded as the fallback radius. In addition to the three discontinuities, there is an interface where the flow velocity changes its sign, i.e., the in- and outflow boundary. When $r_{\mathrm{fb}}$ is decreasing, the in- and outflow boundary exists between the contact surface and the reverse shock. On the other hand, when $r_{\mathrm{fb}}$ is increasing, the in- and outflow boundary exists between the contact surface and the forward shock.

\begin{figure}[htp]
    \centering
    \includegraphics[width=19cm]{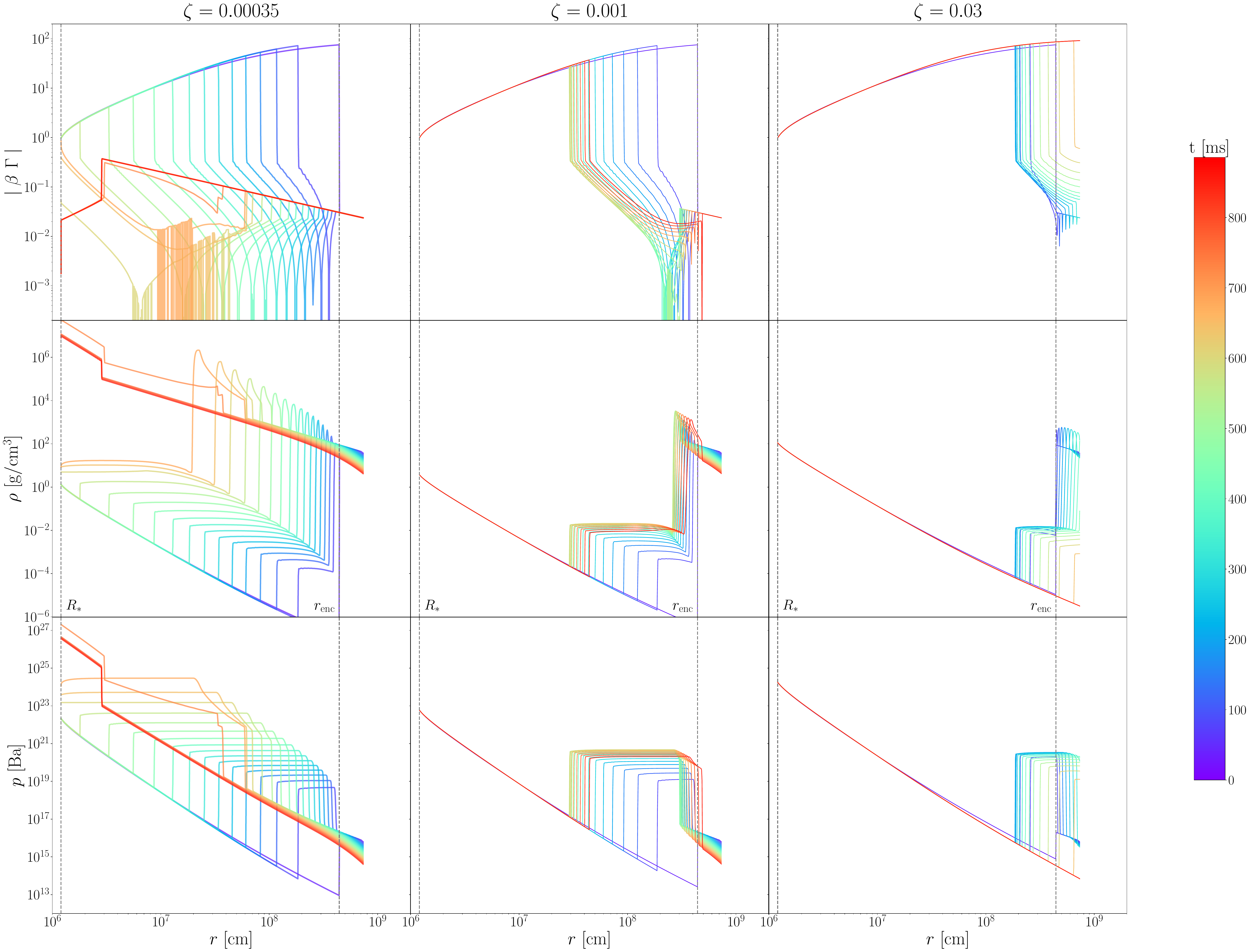}
    \caption{Time evolution of the hydrodynamic structure after a relativistic outflow with a terminal Lorentz factor of $\Gamma_\infty = 100$ collides with a fallback matter at a radius of $r_{\mathrm{enc}} = 4.5 \times 10^{8} $ cm marked by the vertical dashed lines. The cases with three different out- and inflow luminosity ratios are shown : $\zeta=0.00035$ (left), $0.001$ (center) and $0.03$ (right). The top, middle, and bottom panels show the velocity, density, and pressure profiles, respectively, for $0 < t < 885.6\,\mathrm{ms}$.}
    \label{fig:Three discontinuity}
\end{figure}

\begin{figure}[!htp]
    \centering
    \includegraphics[width=18cm]{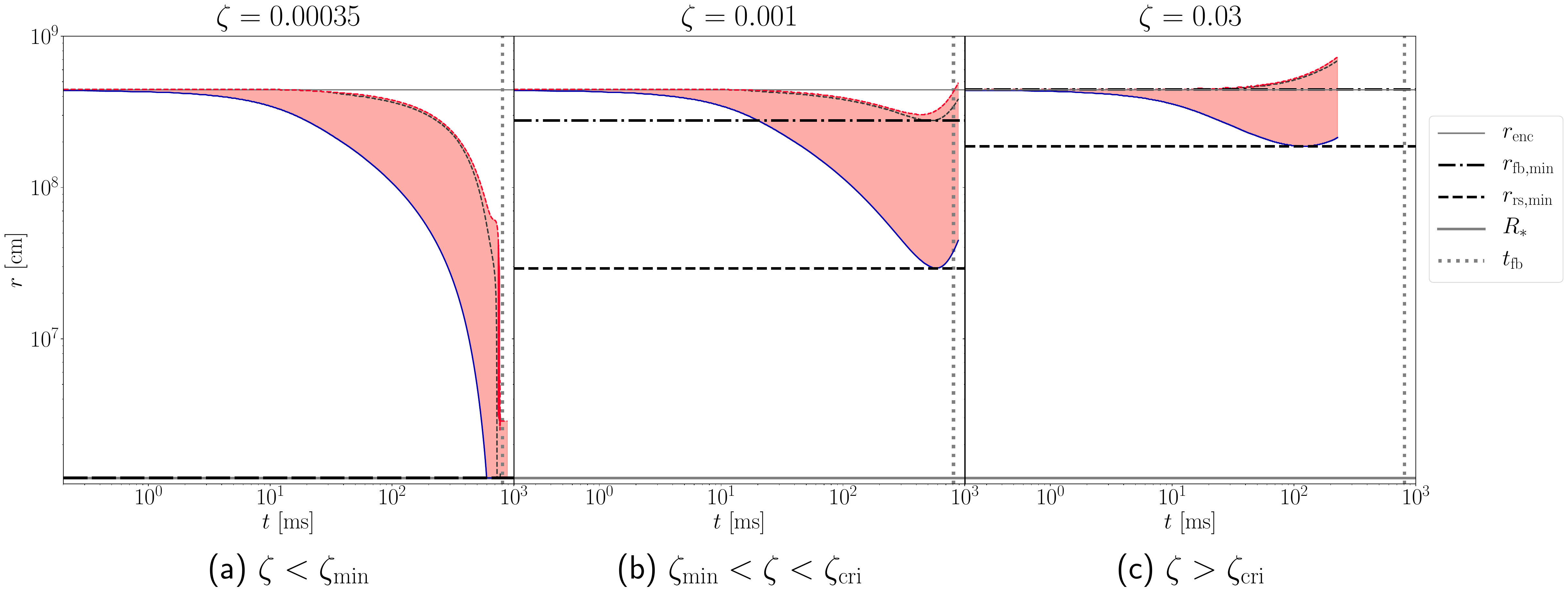}
    \caption{Time evolution of the shock structure of the case shown in Fig. \ref{fig:Three discontinuity}.  The shaded regions indicate the shocked region bounded by the forward shocks (red dashed lines) and reverse shocks (blue solid lines). The encounter radius and NS surface are marked by thin grey and thick grey solid line, respectively. The left panel shows a successful fallback to the NS ($\zeta < \zeta_\mathrm{min}$). The right panel shows a case with $\zeta > \zeta_\mathrm{cri}$, in which the outflow completely repels the fallback matter. The center panel shows an intermediate case with $\zeta_\mathrm{min} < \zeta < \zeta_\mathrm{cri}$, where the matter starts to fall back but becomes overwhelmed by the outflow before it reaches the NS surface.
    }
    \label{fig:innermostradius}
\end{figure}

The time evolution of the shock structure are summarized in Figs. \ref{fig:Three discontinuity} and {\ref{fig:innermostradius}}. 
Fig. \ref{fig:Three discontinuity} shows the velocity (top row), density (middle row), and pressure (bottom row) profile for three cases with the same wind Lorentz factor $\Gamma_\infty = 100$ and encounter radius $r_{\mathrm{enc}} = 4.5 \times 10^{8} $ cm but with different in- and outflow luminosity ratios $\zeta=0.00035$ (left column), $0.001$ (central column) and $0.03$ (right column).  
Fig. {\ref{fig:innermostradius}} shows time evolution of the position of the forward shock, the reverse shock, and the contact surface of the cases shown in Fig. \ref{fig:Three discontinuity}. The shaded regions represent the entire shocked regions. The solid, dash, and dotted-dash horizontal lines indicate the encounter radii, the minimum reverse shock radii $r_\mathrm{rs,min}$, and the minimum fallback radii $r_\mathrm{fb,min}$, respectively, and the vertical dotted line corresponds to $t = t_\mathrm{fb}$.

As shown in Figs. \ref{fig:Three discontinuity} and \ref{fig:innermostradius}, the time evolution of the accretion shock can be basically classified into three types depending on the out- to inflow luminosity ratio $\zeta$. In the small $\zeta$ limit, i.e., the intense fallback limit, the shocked region monotonically contracts (see left panels). The fallback matter reaches to the NS surface in about a free-fall time ($t \sim 600$ ms) from $r = r_\mathrm{enc}$ to $r_\mathrm{rs,min} = r_\mathrm{fb,min} = R_*$.

In the opposite limit, the fallback radius $r_{\mathrm{fb}}$ monotonically increases (see the right panels), where $r_\mathrm{fb,min} = r_\mathrm{enc}$ by definition. 
We note that even in this case the reverse shock radius can decrease for a while after the encounter. 
For the intermediate case, the shocked region initially contracts to the minimum radius and expands afterward.

\begin{figure}[!htp]
    \centering
    \subfigure[$r_{\mathrm{enc}}=4.5 \times 10^{7} $ cm]{
    \label{Fig.xenc_300}
    \includegraphics[width=0.45\textwidth]{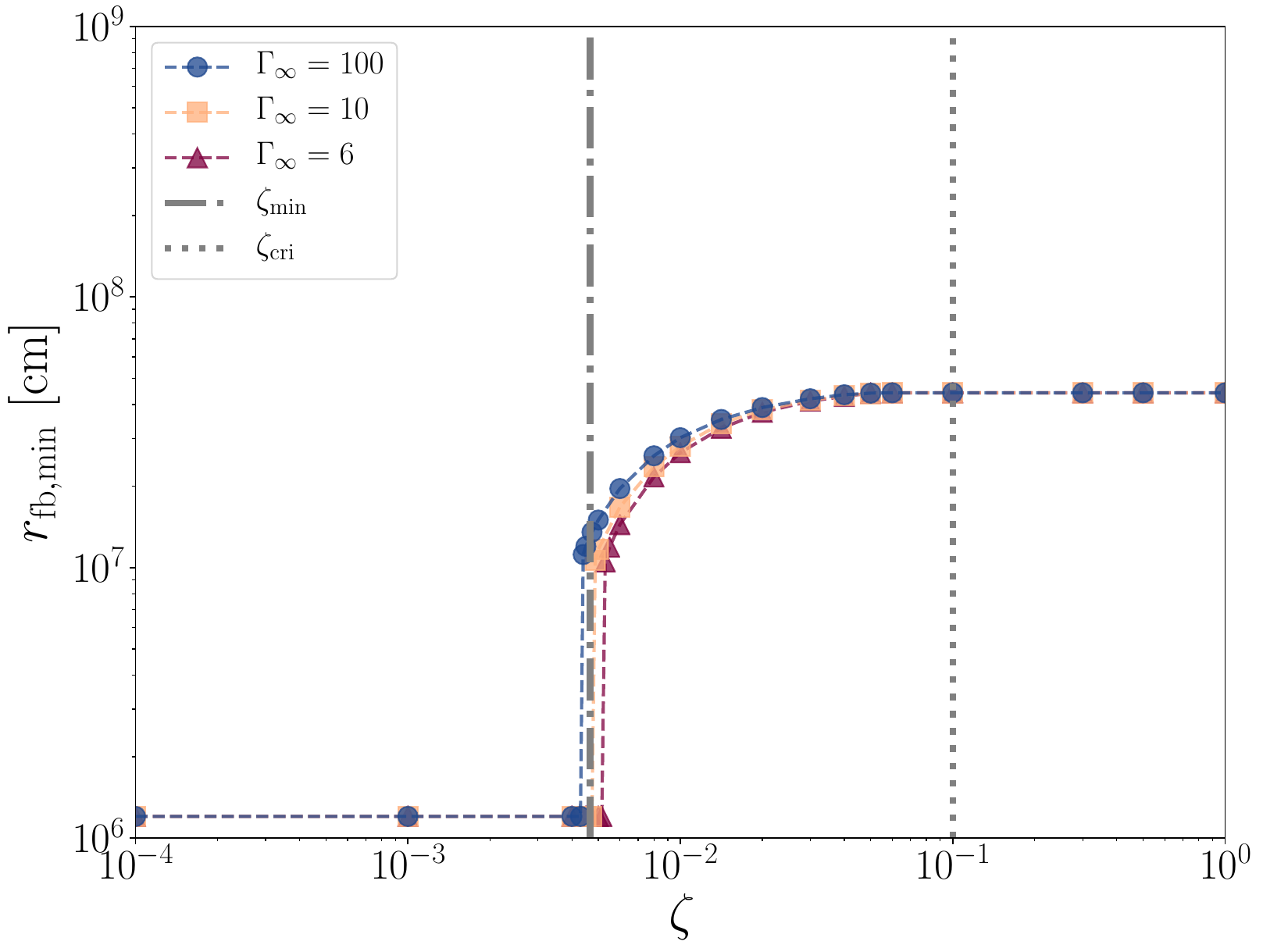}}
    \subfigure[$r_{\mathrm{enc}}=4.5 \times 10^{8} $ cm]{
    \label{Fig.xenc_3000}
    \includegraphics[width=0.45\textwidth]{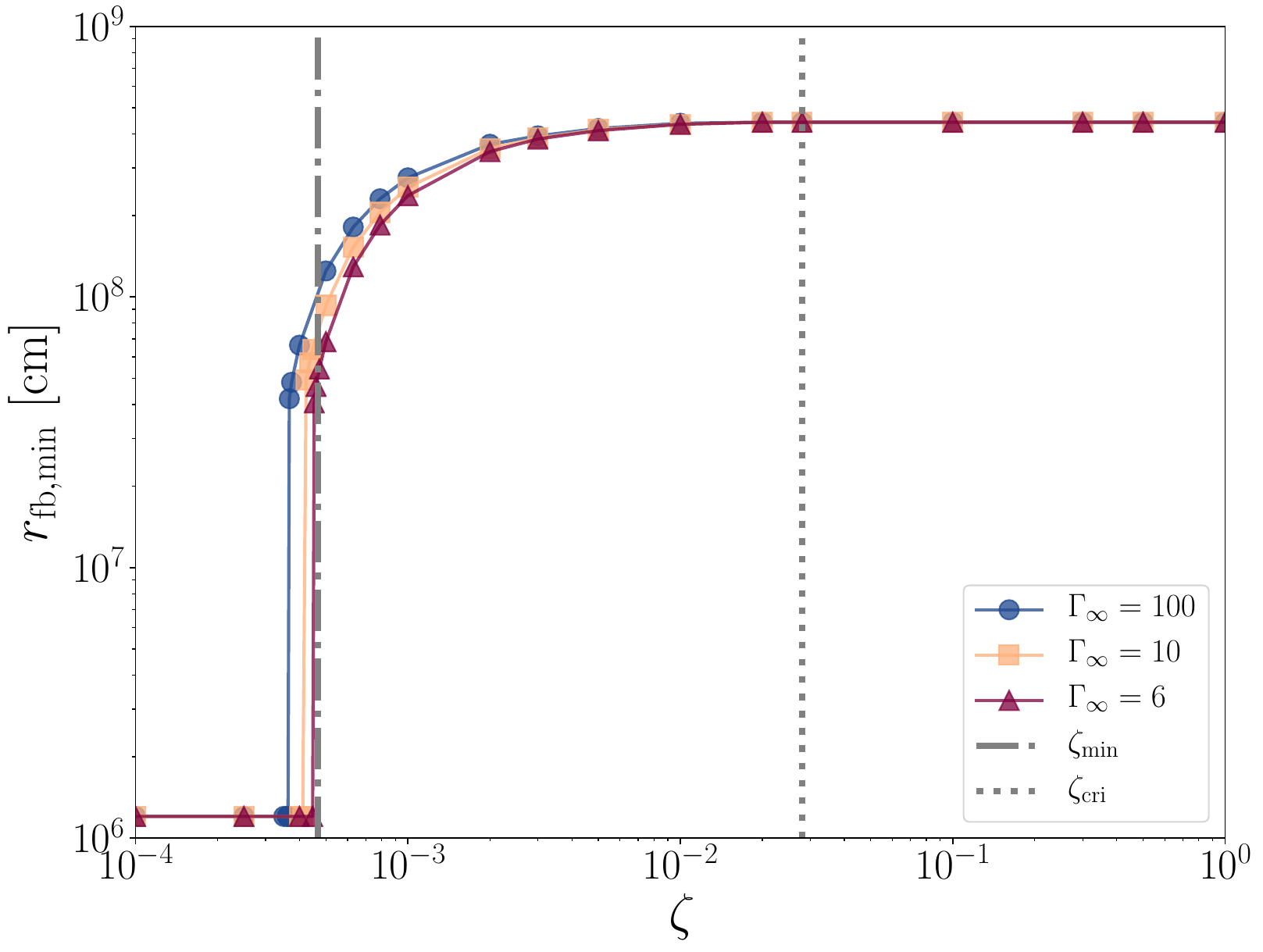}}
    \caption{Dependence of the minimum fallback radius $r_\mathrm{fb, min}$ on the out- to inflow luminosity ratio $\zeta$. The cases with an outflow Lorentz factor of $\Gamma_{\infty}=6$ (wine triangle-up points), $10$ (peach square points) and $100$ (cobalt circle points) at encounter radius of  $r_{\mathrm{enc}}=4.5 \times 10^{7} $ cm (dashed data curve) and $4.5 \times 10^{8}$ cm (solid data curve) are shown. The critical values of $\zeta_\mathrm{cri}$ (Eq. \ref{eq:zeta_cri}) and $\zeta_\mathrm{min}$ (Eq. \ref{eq:zeta_min}) are marked by the vertical dotted and dashed lines, respectively.} 
    \label{fig:before fitting_ft}
\end{figure}

The minimum fallback radius $r_\mathrm{fb, min}$ is of the most important for characterizing its dynamics and determining the fate of the central NS. 
Fig. \ref{fig:before fitting_ft} summarizes our series of simulations, showing the dependence of $r_\mathrm{fb, min}$ on the out- to inflow luminosity ratio $\zeta$ for the cases with two different encounter radii $r_{\mathrm{enc}}=4.5 \times 10^{7}\,\mathrm{cm}$ and $4.5 \times 10^{8}\,\mathrm{cm}$. 
The three different types of the accretion shock dynamics shown in Figs. \ref{fig:Three discontinuity} and \ref{fig:innermostradius} are separated by the two critical values, $\zeta_\mathrm{cri}$ and $\zeta_\mathrm{min}$.
Firstly, $\zeta_\mathrm{cri}$ separates the monotonically expanding cases from the intermediate cases.
For $\zeta > \zeta_{\rm{cri}}$, the ram pressure of the relativistic wind should be larger than that of the fallback inflow at the encounter. Based on this consideration, we can analytically derive $\zeta_\mathrm{cri}$ from the ram pressure balance at the encounter radius
\begin{equation}\label{eq:cond_expand_num}
    \frac{L}{4\pi r_\mathrm{enc}{}^2 c} \gtrsim \frac{\dot M_\mathrm{fb,ini} v_\mathrm{fb}(r_\mathrm{enc})}{4\pi r_\mathrm{enc}{}^2}, 
\end{equation}
or 
\begin{equation}\label{eq:zeta_cri}
    \zeta \gtrsim \zeta_{\rm{cri}} = \left(\frac{2GM_*}{r_\mathrm{enc}c^2}\right)^{1/2} = {\cal R}{\rm_{enc}}^{-1/2}.
\end{equation}
Eq. (\ref{eq:zeta_cri}) is indicated by dotted lines in Fig. \ref{fig:before fitting_ft} and consistent with our numerical results. 

\begin{figure}[!htp]
    \centering
    \includegraphics[width=11cm]{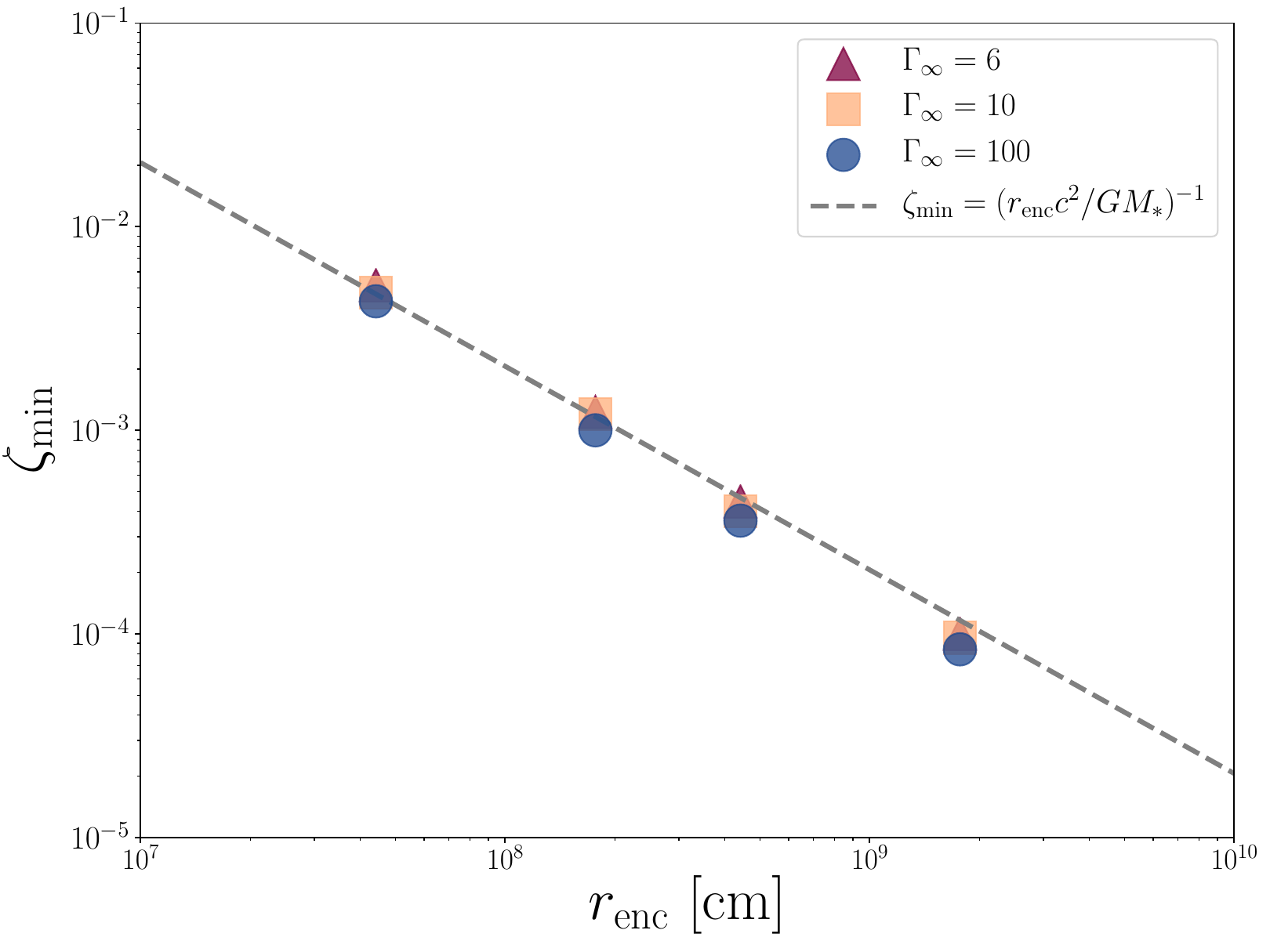}
    \caption{Dependence of the largest out-inflow luminosity ratio for the fallback to invade down to the NS surface $\zeta_\mathrm{min}$ on the encounter radius. For $\zeta < \zeta_\mathrm{min}$, the fallback matter reaches to the NS surface. The cases with wind Lorentz factor of $\Gamma_{\infty}=6$ (wine triangle-up points), $10$ (peach square points) and $100$ (cobalt circle points), respectively. The dashed line corresponds to Eq. (\ref{eq:zeta_min}).} 
    \label{fig:zeta_minVSrenc}
\end{figure}

For $\zeta < \zeta_\mathrm{cri}$, the minimum fallback radius decreases as $\zeta$ decreases. In particular, it exponentially decreases  at around another critical value $\zeta_\mathrm{min}$, and $r_\mathrm{fb, min} = R_*$ for $\zeta < \zeta_\mathrm{min}$. 
Fig. \ref{fig:zeta_minVSrenc} summarizes the dependence of this critical value $\zeta_\mathrm{min}$ with respect to the encounter radius and outflow Lorentz factor. 
We find that $\zeta_\mathrm{min}$ is inversely proportional to $r_\mathrm{enc}$ as
\begin{equation}\label{eq:zeta_min}
\zeta_{\min } \approx  \frac{G M_*}{c^2 r_{\mathrm{enc}}}.
\end{equation}
This result can be interpreted in light of a simplified thin-shell model (Appendix \ref{app:thinshell}), 

in which we approximate the shocked matter as a shell at the contact surface $r_\mathrm{fb}$, and its dynamics can be obtained through solving simplified mass, momentum, and energy conservation equations including the effects of gravity. 
We find that $\zeta \approx \zeta_\mathrm{min}$ corresponds to the case where the time-integrated outflow luminosity and the gravitational work exerted to the shocked fallback matter become comparable at $t \approx t_\mathrm{fb}$. In this case, the outflow can marginally repel the fallback matter. For a smaller $\zeta < \zeta_\mathrm{min}$, the outflow cannot supply a sufficient amount of energy to the shocked region by the time gravity accelerates the fallback.
We also note that Eq. (\ref{eq:zeta_min}) is broadly consistent with the minimum out- to inflow luminosity ratio for the existence of a self-similar solution describing the expanding accretion shock~\footnote{See \cite{SK18} and their Eqs. 31 and 32, where the dimensionless out- to inflow luminosity ratio is defined as $4\pi D_\mathrm{fs} \sqrt{\xi_\mathrm{s}}$ in their Eq. (27).}.

In Figs. \ref{fig:before fitting_ft}  and \ref{fig:zeta_minVSrenc}, we also show the cases with three different outflow Lorentz factors $\Gamma_{\infty} = 6$, $10$, and $100$. It is found that $r_\mathrm{fb, min}$ and so as $\zeta_\mathrm{min}$ barely change with $\Gamma_{\infty}$~\footnote{We also confirm that the results hold for a mildly relativistic case with $\Gamma_{\infty} = 2$.}. As argued in the previous paragraph, $r_\mathrm{fb, min}$ is determined by the balance between the time-integrated outflow luminosity injected to and the gravitational work exerted to the shocked matter, neither of which depends on the outflow velocity as long as it is relativistic. Thus, although we only explored the cases with $\Gamma_\infty \leq 100$, Eq. (\ref{eq:zeta_min}) can be applicable to cases with a larger outflow Lorentz factor. 

We note that the fallback radius in the contracting phase will be subject to the Rayleigh-Taylor (RT) instability given the velocity, density and pressure profiles. When the RT instability is induced, the so-called RT fingers will be developed and the fallback accretion will break spherical symmetry. 
We will investigate the impacts of the instability, in particular on the critical condition (Eq. \ref{eq:zeta_min}) in future works.

\section{Implications for the diversity in young neutron stars}\label{sec:implication}

In the previous section, we derive a necessary condition for supernova fallback confronting with a relativistic outflow to reach the near NS surface, i.e., $\zeta < \zeta_\mathrm{min}$. By assuming that the relativistic outflow and fallback matter typically encounters at the initial fallback radius (Eq. \ref{eq:r_enc}), the critical condition can be described in terms of the outflow luminosity $L$, the fallback mass $M_\mathrm{fb}$, and the fallback time $t_\mathrm{fb}$ as 
\begin{equation}\label{eq:Mfb_con}
    M_\mathrm{fb, crit} \approx \frac{5}{2} \times (G M_*)^{-2/3} L t_\mathrm{fb}{}^{5/3}.
\end{equation}
In general, $L$ depends on the rotation period, the strength and configuration of the surface magnetic field. 

While the magnetosphere of the NS is not disturbed by the fallback accretion, the spindown luminosity can be approximated by the dipole formula; 
\begin{equation}\label{eq:Ld}
    L_\mathrm{d} = \frac{B_*{}^2 \Omega_\mathrm{i}{}^4 R_*^6}{4c^3} (1+\sin \chi^2) \sim 4.3 \times 10^{41}\,{\rm erg\,s^{-1}}\,(1+\sin \chi^2)\,B_{*, 13}{}^2 P_{\mathrm{i},-2}{}^{-4},
\end{equation}
with $B_{\rm *}$ being the surface field strength, $\Omega_\mathrm{i} = 2\pi/P_\mathrm{i}$ being the initial angular frequency, and $R_*$ being the NS radius; $\chi$ is the inclination angle between the rotation and dipole axes~\citep{Gruzinov05,Spitkovsky06,Tchekhovskoy+13}.
The spindown timescale can thus be estimated as
\begin{equation}\label{eq:tsd}
    t_\mathrm{sd} \sim 23.5\,\mathrm{yr}\,(1+\sin \chi^2)^{-1}B_{*, 13}{}^{-2}P_{\mathrm{i},-2}{}^{2}.
\end{equation}

Substituting Eq. (\ref{eq:Ld}) to Eq. (\ref{eq:Mfb_con}), 
\begin{equation}\label{eq:Mfb_con_d}
    M_\mathrm{fb, crit} \sim  7.7\times 10^{-8}\,M_\odot\,(1+\sin \chi^2)\,B_{*, 13}{}^2 P_{\mathrm{i},-2}{}^{-4}t_\mathrm{fb, 1}{}^{5/3}  \ \ \text{(dipole)}.
\end{equation}
If the fallback mass is smaller than $M_\mathrm{fb, crit} $, the fallback matter is repelled by the dipole spin-down power. 
Otherwise, the fallback continues as can be seen in the left panel of Fig. \ref{fig:Three discontinuity}. As mentioned Sec. \ref{numerical_simu}, the contact surface is subject to the RT instability, and the fallback proceeds in an anisotropic manner. When the most advanced channeled flow reaches the 
near NS surface, it compresses the magnetosphere down to the size of the Alfv$\acute{\text{e}}$n radius; 
\begin{equation} \label{eq_rm}
    r_\mathrm{A} = \left(\frac{B_*{}^2R_*{}^6}{\dot M_\mathrm{fb} \sqrt{2GM_*}}\right)^{2/7} \sim 1.1\times10^{6}\,\mathrm{cm}\,B_{*, 13}{}^{4/7} M_\mathrm{fb, -4}{}^{-2/7} t_\mathrm{fb, 1}{}^{2/7}.
\end{equation}
Note that, in the cases of our interest, the Alfv$\acute{\text{e}}$n radius is basically smaller than the light cylinder (Eq. \ref{r_lc}) and the corotation radius
\begin{equation}
    r_\mathrm{co} = \left(\frac{GM_*}{\Omega_\mathrm{i}{}^2}\right)^{1/3} \sim 7.8\times 10^{6}\,\mathrm{cm}\,P_\mathrm{i, -2}{}^{2/3}. 
\end{equation}
Such an accretion can expand the polar cap region of open magnetic field lines and enhance the spindown torque of the NS~\citep[e.g.,][]{2016ApJ...822...33P,2018ApJ...857...95M}. 
In this case, the spin-down power can be described as 
\begin{equation}\label{eq:Lm}
    L_\mathrm{m} \approx \begin{cases} 
    (B_*{}^2 \Omega_\mathrm{i}{}^4 R_*^6/c^{3}) \times (r_\mathrm{lc}/r_\mathrm{m})^2\sim 3.1 \times 10^{45}\,{\rm erg\,s^{-1}}\,B_{*, 13}{}^{6/7} P_{\mathrm{i},-2}{}^{-2}M_\mathrm{fb, -4}{}^{4/7} t_\mathrm{fb, 1}{}^{-4/7} & r_\mathrm{A} > R_\mathrm{*} \\
    (B_*{}^2 \Omega_\mathrm{i}{}^4 R_*^6/c^{3}) \times (r_\mathrm{lc}/R_\mathrm{*})^2 \sim 2.7 \times 10^{45}\,{\rm erg\,s^{-1}}\,B_{*, 13}{}^2 P_{\mathrm{i},-2}{}^{-2} & r_\mathrm{A} \leq R_\mathrm{*}
    \end{cases}. 
\end{equation}
The latter case corresponds to the split monopole configuration, that yields the possible maximum power for a given set of $B_*$ and $P_\mathrm{i}$.
The bulk of the fallback matter except for those accreted through the advanced channeled flows will confront with this enhanced outflow. Note that the spin-down power will change with either the spin-down timescale or the accretion timescale. The both are at least comparable to the overall fallback timescale, thus the luminosity can be approximated as constant. 
Substituting Eq. (\ref{eq:Lm}) to Eq. (\ref{eq:Mfb_con}), the critical condition is given as 
\begin{equation}\label{eq:Mfb_con_m}
    M_\mathrm{fb, crit} \sim \begin{cases}
    5.2\times10^{-3}\,M_\odot\,B_{*, 13}{}^{2} P_{\mathrm{i},-2}{}^{-14/3} t_\mathrm{fb, 1}{}^{23/9} & r_\mathrm{A} > R_\mathrm{*} \\
    4.8\times10^{-4}\,M_\odot\,B_{*, 13}{}^2 P_{\mathrm{i},-2}{}^{-2}t_\mathrm{fb, 1}{}^{5/3} & r_\mathrm{A} \leq R_\mathrm{*} 
    \end{cases}.
\end{equation}
If the fallback mass is smaller than Eq. (\ref{eq:Mfb_con_m}), the fallback matter is repelled by the enhanced spin-down power. Otherwise, the bulk of the fallback matter reaches the near surface region, and the newly formed magnetosphere is expected to be strongly disturbed. In particular, if the fallback mass is larger than Eq. (\ref{eq:Mfb_con_m}) and $r_A \leq R_*$, or 
\begin{equation}\label{eq:Mfb_con_b}
    M_\mathrm{fb} > 8.2\times 10^{-5}\,M_\odot\,B_{*, 13}{}^{2} t_\mathrm{fb, 1},
\end{equation}
the fallback matter can enshroud and bury the surface magnetic fields.  
\begin{figure}[!htp]
    \centering
    \includegraphics[width=0.55\textwidth]{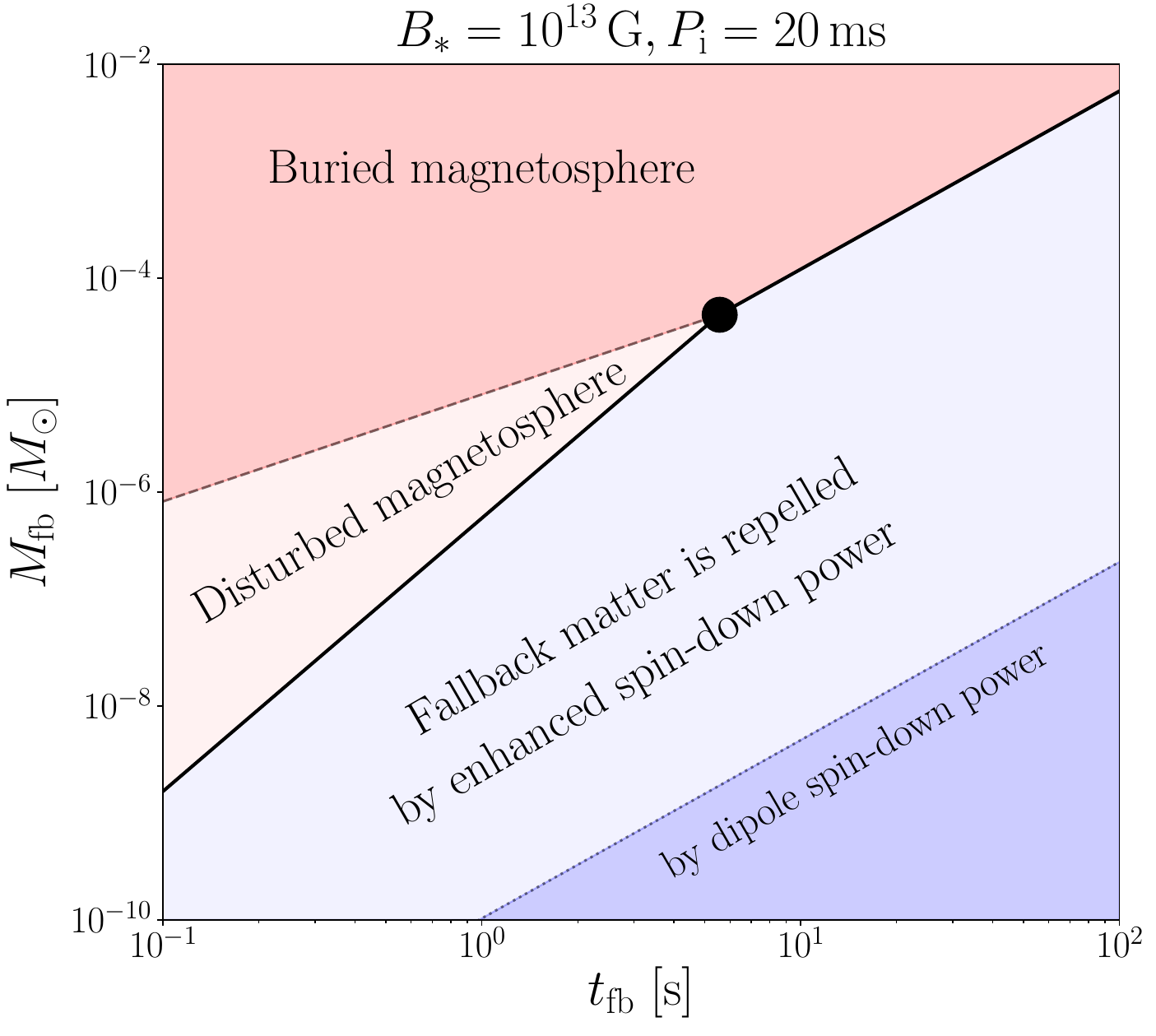}
    \caption{Possible consequences of the collision between rotation-powered wind from a newborn neutron star with a surface magnetic field $B_* = 10^{13}\,\mathrm{G}$ and an initial spin period $P_\mathrm{i} = 20\,\mathrm{ms}$ and supernova fallback with total fallback mass $M_\mathrm{fb}$ and fallabck timescale $t_\mathrm{fb}$.}
    \label{fig:crab}
\end{figure}

\begin{figure}[!htp]
    \centering
    \includegraphics[width=1\textwidth]{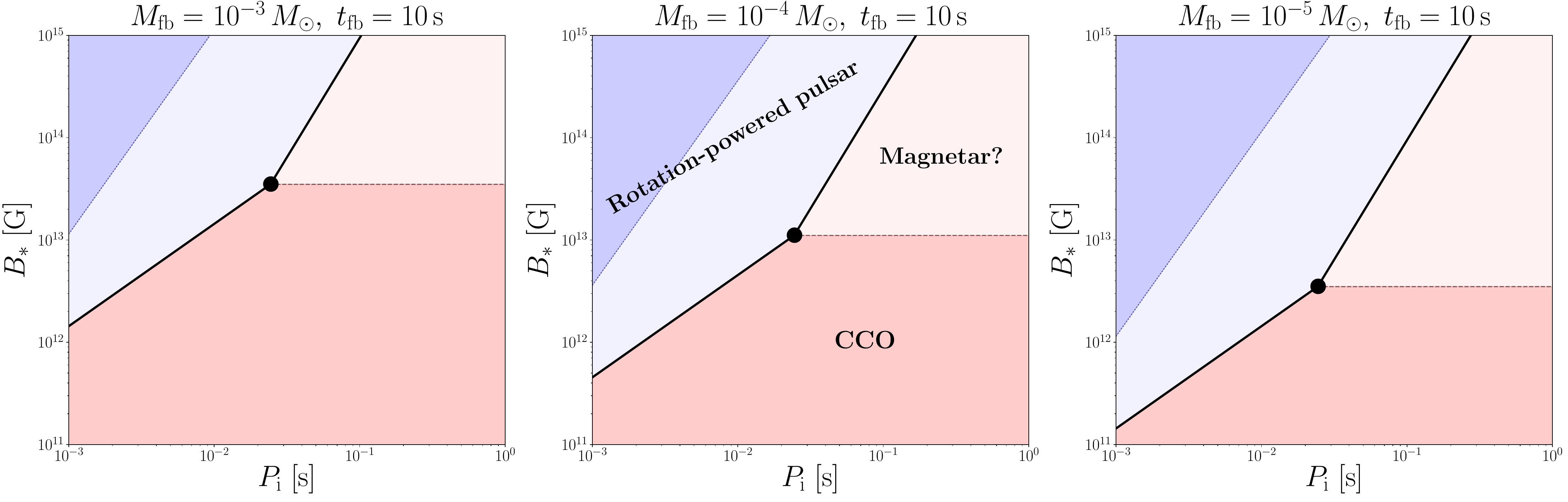}
    \caption{Possible trifurcation in the types of neutron star caused by the interaction between the rotation-powered wind and the supernova fallback in the newborn phase. The left, center, right panels show the cases with fixed fallback time $t_\mathrm{fb} = 10\,\mathrm{s}$ and fallback mass of $M_\mathrm{fb} = 10^{-3}\,M_\odot$, $10^{-4}\,M_\odot$, $10^{-5}\,M_\odot$, respectively.}
    \label{fig:diversity}
\end{figure}

Fig. \ref{fig:crab} summarizes the above discussions: the dotted line indicates the condition Eq. (\ref{eq:Mfb_con_d}), the solid line corresponds to the condition Eq. (\ref{eq:Mfb_con_m}), and the dashed line shows the boundary set by the condition Eq. (\ref{eq:Mfb_con_b}).
\begin{itemize}
\item For cases below the solid line, the fallback matter is repelled either by the dipole or enhanced spin-down power. A force-free magnetosphere will be restored even for the latter case after the channel flow to the pole region ceases. These NSs can naturally evolve into rotation-powered pulsars. 
\item For cases above the dashed line, the surface magnetic fields can be buried down in the outer crust, then the apparent magnetic field strength becomes significantly weaker. These NSs can be observed as CCOs. 
\item The final cases are those above the solid line but below the dashed line. The fallback accretion is intense enough for reaching the NS surface, but not intense enough for burying the surface fields. 
In this case, the magnetosphere will be strongly disturbed in a chaotic manner for a fallback timescale. Although to address the detailed field configuration of the resultant magnetosphere is beyond the scope of this paper, we speculate that the quasi-spherical compression of the rotating magnetosphere could result in synthesizing multipolar surface magnetic fields with an enhanced field strength. In addition to the internal amplification of the magnetic field in the core-collapse and the proto-NS phase, such an external amplification by the fallback accretion might be important to form magnetars.
\end{itemize}

In the proposed scenario, the branching into three different NS populations occurs at the intersection of the solid and dashed lines in Fig. \ref{fig:crab} marked by a black point; 
\begin{equation}\label{eq:B_bif}
    B_\mathrm{*, tri} \sim 1.1 \times 10^{13}\,\mathrm{G}\, M_\mathrm{fb, -4}{}^{1/2}t_\mathrm{fb, 1}{}^{-1/2}.
\end{equation}
\begin{equation}\label{eq:P_bif}
    P_\mathrm{i, tri} \sim 24\,\mathrm{ms}\,t_\mathrm{fb, 1}^{1/3}.
\end{equation}
Note that $P_\mathrm{i, tri}$ does not depend on the fallback mass.  
For a typical range of the fallback accretion with $M_\mathrm{fb} \sim 10^{-2}\mbox{-}10^{-4}\,M_\odot$ and $t_\mathrm{fb} \sim 1\mbox{-}100\,\mathrm{s}$, Eqs. (\ref{eq:B_bif}) and (\ref{eq:P_bif}) imply that the trifurcation occurs at $B_\mathrm{*} \sim 10^{13}\,\mathrm{G}$ and $P_\mathrm{i} =$ a few 10 ms (see Fig. \ref{fig:diversity}). Such magnetic field strength and rotation period at birth are broadly consistent with those inferred for Galactic rotation-powered pulsars, i.e., a typical pulsar formation occurs at around the triple point. This can naturally explain the observed fact that the formation rate of rotation-powered pulsars is roughly comparable to those of CCOs and magnetars~\citep[e.g.,][]{2008MNRAS.391.2009K}.

\if
Here we obtain the complete fitting formula for each $\zeta$ vs $r_\mathrm{fb,min}$ shown in Fig. \ref{fig:before fitting_ft} (see Appendix, Fig. \ref{fig:after fitting}), and if we substitute the minimum fallback radii as the light cylinder radius $r_{lc}$ shown in Eq.(\ref{r_lc}) and assume the dipole magnetic field configuration (see Eq.(\ref{eq:Ld})) since the accretion never invade down to the NS surface at this moment, the criterion for the fallback to reach the magnetosphere will be given in Eq.(\ref{rl_criterion}), which may give hints for future MHD study.
\begin{equation}
\frac{r_{\mathrm{fb}, \min }}{r_{\mathrm{enc}}}=\frac{a \log (\zeta)+b}{\log (\zeta)+d} < \frac{r_{lc}}{r_{\mathrm{enc}}} = 0.1 \times \Omega_\mathrm{i,2}{}^{-1} t_{\rm fb, 1}{}^{-2/3} \equiv k_{\Omega,t},
\end{equation}
\begin{equation}
\zeta = \frac{L_\mathrm{d}}{\dot M_\mathrm{fb,ini}c^2} < e^{(dk_{\Omega,t}-b)/(a-k_{\Omega,t})},
\end{equation}
\begin{equation} \label{rl_criterion}
\dot M_\mathrm{fb,ini} > \dot M_\mathrm{fb,rl} = 1.54 \times 10^{-15}\,(1+\sin \chi^2)\,B_{\rm d, 13}{}^2 \Omega_{\mathrm{i},2}{}^4 e^{(b-dk_{\Omega,t})/(a-k_{\Omega,t})}\,M_\odot\,/s
\end{equation}

If the fallback accretion rate $\dot{M}_\mathrm{fb,i}$ keep increasing based on Eq.(\ref{rl_criterion}), it is possible for the accretion to reach the NS surface, which is characterized by the critical out-inflow luminosity ratio for the fallback to be marginally repelled by the newborn pulsar wind $\zeta_\mathrm{min}$: the fallback matter will be able to reach the NS surface while $\zeta < \zeta_\mathrm{min}$ (or $\dot{M}_\mathrm{fb,i} > \dot{M}_\mathrm{cri,repel}$). At this moment, one shall consider the fact that the surface magnetic field might be maximumly open due to the fallback accretion, which will lead to the change of magnetic field configuration from dipole to split monopole-like, as shown in Eq.(\ref{eq:Lm}). If one assume that the spin-down luminosity under normal dipole and split monopole-like magnetic field configuration, the criterion for the fallback to be marginally repelled by the new-born pulsar wind can be written as Eq.(\ref{eq:repel_d}) and Eq.(\ref{eq:repel_m}), respectively.

\begin{equation} \label{eq:repel_d}
\zeta = \frac{L_\mathrm{d}}{\dot M_\mathrm{fb,ini}c^2} < \zeta_\mathrm{min} = \frac{L_\mathrm{d}}{\dot M_\mathrm{cri,repel}c^2} \approx \left(\frac{ r_{\mathrm{enc}} c^{2}}{G M_*}\right)^{-1},
\end{equation}
 
\begin{equation} \label{eq:repel_m}
\zeta = \frac{L_\mathrm{m}}{\dot M_\mathrm{fb,ini}c^2} < \zeta_\mathrm{min} = \frac{L_\mathrm{m}}{\dot M_\mathrm{cri,repel}c^2} \approx \left(\frac{ r_{\mathrm{enc}} c^{2}}{G M_*}\right)^{-1},
\end{equation}

With showing the contour of $\dot{M}_\mathrm{cri,repel}$ for both monopole-like (see solid lines in the right panel) and dipole (see dashed lines in the left panel) magnetic field configuration regarding $\{B_{\rm *},P_{\rm i}\}$ (Characteristic magnetic field and rotation period) of central NS in Fig. \ref{fig:Bd_Pi} assuming fallback time scale $t_\mathrm{fb}=10$ s, the parameter space that fallback with certain mass accretion rate $\dot{M}_\mathrm{fb,i}$ will be repelled by pulsar wind from NS with split monopole-like and dipole magnetic field configuration can be shown as the region above the dashed and solid contour corresponding to the value of $\dot{M}_\mathrm{fb,i}$, respectively.

Eq.(\ref{bury_criterion}) can potentially be applied to the Crab-like pulsar as well. Crab-like pulsar is a young radio pulsar with rotation period $P_{i} = 33.1$ ms and characteristic magnetic field $B_{*} = 7.6 \times 10^{12}$ G. Assuming that the inclination angle $\chi=0$, one could obtain the dependence of total accretion mass $M_\mathrm{fb}$ that the fallback matter could marginally invade down to the NS surface with assuming dipole magnetic field configuration (see cobalt solid line) and monopole-like magnetic field configuration (see pink solid line) on the fallback time scale $t_{\mathrm{fb}}$ in Fig. \ref{fig:Mfb_tfb}, respectively. From which we could conclude that a pulsar will be largely expected in the region where the fallback accretion rate $\dot{M}_\mathrm{fb,i}<\dot{M}_\mathrm{cri,repel}$.

\fi

\section{Summary and discussion}\label{sec:discussion}
By performing a set of relativistic hydrodynamic simulations, we investigate the accretion shock formed between supernova fallback matter and confronting relativistic outflow. 
We find that the time evolution of the accretion shock can be basically classified into three types depending on the encounter radius of the flows $r_\mathrm{enc}$ and a dimensionless parameter $\zeta \equiv L/\dot M_\mathrm{fb}c^2$. 
The accretion shock monotonically expands when $\zeta \gtrsim \zeta_\mathrm{cri} \equiv (2GM_*/c^2r_\mathrm{enc})^{1/2}$ while monotonically contracts and reaches the stellar surface when $\zeta \lesssim \zeta_\mathrm{min} \equiv GM_*/c^2r_\mathrm{enc}$, where $M_*$ is the NS mass. For the intermediate cases ($\zeta_\mathrm{min} \lesssim \zeta \lesssim \zeta_\mathrm{cri}$), the accretion shock initially contracts but start to expand before reaching the surface. 
We confirm that the results are not sensitive to the Lorentz factor of the wind. 

Based on the results, we discuss the possible consequences of supernova fallback on nascent NSs; when the fallback matter is repelled by the spin-down power, the NS successfully evolves to a rotation powered pulsar. Otherwise the fallback accretion invading down to the NS surface strongly compresses the magnetosphere, which is either buried under the outer crust or reconfigured to form enhanced multipolar fields. The former and latter cases may result in forming CCOs and magnetars, respectively. Our calculations suggest that, for a typical range of supernova fallback with $M_\mathrm{fb} \sim 10^{-(2\mbox{-}4)}\,M_\odot$ and $t_\mathrm{fb} \sim 1\mbox{-}100\,\mathrm{s}$, an NS with a magnetic field strength of $B_* \sim 10^{13}$ G and a rotation period of $P_\mathrm{i} =$ a few 10 ms is at the triple point of the three different NS populations. 

We note that our numerical results are obtained by spherically symmetric relativistic hydrodynamic simulations. The fallback accretion and the relativistic outflow from the nascent NS are in general anisotropic, and the contact surface between the in- and outflows will be subject to the RT instability, the consequences of which cannot be captured by our one-dimensional study. Multi-dimensional simulations are desirable for quantifying the impacts of such effects on the critical fallback condition. 
In addition, magnetohydrodynamics simulations are also important especially for the cases where the fallback matter invading down to the magnetosphere; whether and how the magnetosphere is reconfigured and/or buried by the fallback matter? These points will be investigated in our future work.

\acknowledgements
The authors thank Naoki Yoshida and Tilman Hartwig for fruitful discussions and technical support.
YZ is supported by the International Graduate Program for Excellence in Earth-Space Science (IGPEES) at the University of Tokyo. This work is also supported by JSPS KAKENHI Grant Numbers JP16H06341, JP20H05639, JP20K04010, JP20H01904, JP18K13579, MEXT, Japan.

\appendix

\section{Thin shell model for the shocked fallback matter}\label{app:thinshell}

We here construct a simplified thin-shell model describing the dynamics of supernova fallback confronting with an energy injection from the central source, in order to better interpret the numerical results presented in Sec. \ref{sec:result}, in particular, the dependence of the minimum fallback radius on the out- to inflow luminosity ratio (Figs. \ref{fig:before fitting_ft} and \ref{fig:zeta_minVSrenc}).

We approximate the shocked fallback matter as a shell at $r = r_\mathrm{fb}$ with a velocity of $v_\mathrm{fb}$ and a mass of $M_\mathrm{fb}$. The mass, momentum, and energy conservation equations can be described as 
\begin{equation}\label{eq:mass_con_thin}
    \frac{dM_\mathrm{fb}}{dt} = -4\pi r_\mathrm{fb}^2 \rho(v - v_\mathrm{fb}),
\end{equation}

\begin{equation}\label{eq:mom_con_thin_grav}
\frac{d (M_{\mathrm{fb}} v_{\mathrm{fb}})}{d t}=4 \pi r_{\mathrm{fb}}^{2} p-\dot M_\mathrm{fb,ini}\left(v-v_{\mathrm{fb}}\right)-\frac{GM_*M_\mathrm{fb}}{r_\mathrm{fb}^2},
\end{equation}

\begin{equation}\label{eq:ene_con_thin}
    3\frac{d}{dt} (pV) + p \frac{dV}{dt} = L,
\end{equation}
Where $v = \sqrt{2GM_*/r_\mathrm{fb}}$, $\rho=\dot M_\mathrm{fb,ini} /\left(4 \pi r_{\mathrm{fb}}^{2} v\right)$ and $V = 4\pi r_\mathrm{fb}^3/3$.
We note that the dynamics of the thin shell can be described by non-relativistic equations because the velocity of the shocked region is well below the speed of light.
Hereafter we assume that the outflow luminosity and mass fallback rate is constant during the evolution.

\subsection{Asymptotic solutions for small $t$} \label{asymptotic_behavior}
Let us first obtain the asymptotic solutions for small $t$. In this case, the velocity, mass, and pressure of the thin shell can be expressed as 

\begin{equation}\label{eq:v_ts}
    v_\mathrm{fb}(t) = v_{\mathrm{fb},0} + at,
\end{equation}
\begin{equation}\label{eq:Mfb_ts}
    M_\mathrm{fb}(t) = \dot{M_\mathrm{s}}t + \ddot{M_\mathrm{s}}t^2/2,
\end{equation}
\begin{equation}\label{eq:p_ts}
    p(t) = L/(4\pi r_\mathrm{enc}^2 c) + \dot{p}t + \ddot{p}t^2/2,
\end{equation}
where $a$ stands for the acceleration of the fallback shell. 
We set the initial conditions as $r_\mathrm{fb,0} = r_\mathrm{enc}$, $v_\mathrm{fb}(t) =v_\mathrm{fb,0}$, $M_\mathrm{fb,0} = 0$, and $4\pi r_\mathrm{enc}^2 p_0 = L/c$.
By substituting Eqs. (\ref{eq:v_ts}), (\ref{eq:Mfb_ts}) and (\ref{eq:p_ts}) into Eqs. (\ref{eq:mass_con_thin}), (\ref{eq:mom_con_thin_grav}) and (\ref{eq:ene_con_thin}), 
one obtains 
\begin{equation}\label{eq:initial_velocity}
v_{\mathrm{fb},0}=\sqrt{\frac{2GM}{r_\mathrm{enc}}}\left(\sqrt{\zeta/\zeta_\mathrm{cri}}-1\right) ,
\end{equation}
\begin{equation}
\dot{M_\mathrm{s}} = \dot M_\mathrm{fb,\ i}\sqrt{\zeta/\zeta_\mathrm{cri}},
\end{equation}
\begin{equation}\label{eq:acceleration}
a=\frac{c^4 \left(5 (1-\zeta/\zeta_\mathrm{cri})+7 \sqrt{\zeta/\zeta_\mathrm{cri}}-7 +2 \sqrt{y\zeta/\zeta_\mathrm{cri} }\right)}{12 {GM} y^2},
\end{equation}
with 
\begin{equation}
y = \frac{c^2r_\mathrm{enc}}{2GM}.
\end{equation}
Eq. (\ref{eq:initial_velocity}) shows that the sign of the initial velocity $v_{\mathrm{fb},0}$ is determined by the ratio $\zeta/\zeta_\mathrm{cri}$ greater or less than unity, which is consistent with the numerical result (Eq. \ref{eq:cond_expand_num}). As long as $\zeta/\zeta_\mathrm{cri}$ is not significantly smaller than unity, the shell reaches the minimum fallback radii $r_\mathrm{fb,min}$ at time $t = t_\mathrm{min}=-v_{\mathrm{fb},0} / a$. In this case, the minimum fallback radius can be estimated as 
\begin{equation}\label{eq:asymptotic_thinshell}
r_\mathrm{fb,min}(x, y)=\frac{2 G My}{c^{2}} \times\left[\frac{8\left(\zeta/\zeta_\mathrm{cri}-1\right)-13 \sqrt{\zeta/\zeta_\mathrm{cri}}+13 -2 \sqrt{y\zeta/\zeta_\mathrm{cri} }}{5 \left(\zeta/\zeta_\mathrm{cri}-1\right)-7 \sqrt{\zeta/\zeta_\mathrm{cri}}+7 -2 \sqrt{y\zeta/\zeta_\mathrm{cri} }}\right]. 
\end{equation}
We confirm that Eq. (\ref{eq:asymptotic_thinshell}) is also consistent with the numerically obtained $r_\mathrm{fb, min}$ for $\zeta/\zeta_\mathrm{cri} \sim 1$.

\subsection{Long-term behavior}\label{app:ltb}

Next, let us derive a formal solution applicable to a relatively large $t$. If $\zeta < \zeta_\mathrm{cri}$, the thin shell initially contracts. 
When it reaches the innermost radius $r_\mathrm{fb, min}$ at $t = t_\mathrm{min}$, $v_\mathrm{fb}=0$. In the critical case where the thin shell marginally become gravitationally unbound, $dv_\mathrm{fb}/dt = 0$ should be also realized at $t = t_\mathrm{min}$. By substituting these conditions into Eqs. (\ref{eq:mass_con_thin}) and (\ref{eq:mom_con_thin_grav}), one obtains

\begin{equation}\label{eqq:equili_criterion}
    4 \pi r_\mathrm{fb, min}^2 p = \dot{M}_\mathrm{fb,i}\sqrt{2GM/r_\mathrm{fb, min}} + \frac{GM_*M_\mathrm{fb}}{r_\mathrm{fb, min}^2}.
\end{equation}
The pressure term in the left hand side can be evaluated by integrating Eq. (\ref{eq:ene_con_thin}) over time as
\begin{equation}\label{eq:P_thin_integ}
    r_\mathrm{fb, min}^4 p = r_\mathrm{enc}^4 p_0 + \frac{L}{4 \pi} \int^{t_\mathrm{min}}_0 dt' r_\mathrm{fb}(t').
\end{equation}
By substituting Eq. (\ref{eq:P_thin_integ}) to Eq. (\ref{eqq:equili_criterion}), we can describe the minimum fallback radius as
\begin{equation}\label{eqq:analytical_model}
    r_\mathrm{fb, min} = r_\mathrm{enc} [ \zeta f(t_\mathrm{min})-g(t_\mathrm{min})]^{2/3}  \left(\frac{c^2r_\mathrm{enc}}{2GM}\right)^{1/3},
\end{equation}
where 
\begin{equation}\label{eq:ft}
    f(t_\mathrm{min}) = 1 + \frac{c\int^{t_\mathrm{min}}_0 dt' r_\mathrm{fb}(t')}{r_\mathrm{enc}^2},
\end{equation}
\begin{equation} \label{eq:gt}
g(t_\mathrm{min}) = \frac{G M_{*} M_\mathrm{fb}(t_\mathrm{min})}{c r_\mathrm{e n c}^{2} \dot{M}_\mathrm{fb,i}} 
\end{equation}
The exact values of $f(t_\mathrm{min})$, $g(t_\mathrm{min})$, and $r_\mathrm{fb, min}$ can only be obtained by directly solving Eqs. (\ref{eq:mass_con_thin}), (\ref{eq:mom_con_thin_grav}), and (\ref{eq:ene_con_thin}). However, as long as $t_\mathrm{min} \lesssim t_\mathrm{fb}$, they can be approximated as $f (t_\mathrm{min}) \approx c t_\mathrm{fb}/r_\mathrm{enc}$ and $g(t_\mathrm{min}) \approx GM_* t_\mathrm{fb}/cr_\mathrm{enc}{}^2$. Then, from Eq. (\ref{eqq:analytical_model}), the critical out- to inflow luminosity ratio that gives $r_\mathrm{fb, min} \rightarrow 0$ is roughly estimated as $\zeta_\mathrm{min} \approx g(t_\mathrm{min})/f(t_\mathrm{min}) \approx GM_*/c^2 r_\mathrm{enc}$, which is consistent with the numerical results. 
This can be interpreted as follows; $\zeta f(t)$ and $g(t)$ represent the time-integrated outflow luminosity injected to the thin shell and the work exerted by the gravitational force to the thin shell, respectively, and $\zeta \approx \zeta_\mathrm{min}$ corresponds to the case where these two become comparable at $t \approx t_\mathrm{fb}$.

\bibliography{main}{}

\end{document}